\documentclass[twocolumn,preprintnumbers,amsmath,amssymb,superscriptaddress]{revtex4}

\usepackage{gensymb}
\usepackage{amsmath, amssymb, latexsym}
\usepackage{epsf}
\usepackage{graphicx}
\usepackage{sidecap}
\usepackage{array}
\usepackage{amsmath}
\usepackage{amssymb}
\usepackage{dcolumn}
\usepackage{epstopdf}
\usepackage{bm}

\usepackage{color}

\def \beq {\begin{equation}}
\def \eeq {\end{equation}}
\newcommand{\ca}{CaSn$_{3}$}

\begin{document}
\title{{Fermi surfaces of the topological semimetal {\ca} probed through de Haas van Alphen oscillations}}

\author{K A M Hasan~Siddiquee}
\author{Riffat Munir}
\author{Charuni Dissanayake}
\affiliation {Department of Physics, University of Central Florida, Orlando, Florida 32816, USA}
\author{Xinzhe Hu}
\author{Swapnil Yadav}
\author{Yasumasa Takano} 
\affiliation {Department of Physics, University of Florida, Gainesville, Florida 32611, USA}
\author{Eun Sang Choi}
\affiliation {National High Magnetic Field Laboratory, Florida State University, Tallahassee, Florida 32816, USA}
\author{Duy Le}
\author{Talat S Rahman}
\author{Yasuyuki Nakajima$^{\ast}$}
\affiliation {Department of Physics, University of Central Florida, Orlando, Florida 32816, USA}

\date{\today}
%\pacs{}
\begin{abstract}
\noindent
{In the search of topological superconductors, nailing down the Fermiology of the normal state is as crucial a prerequisite as unraveling the superconducting pairing symmetry. In particular, the number of time-reversal-invariant momenta in the Brillouin zone enclosed by Fermi surfaces is closely linked to the topological class of time-reversal-invariant systems, and can experimentally be investigated. We report here a detailed study of de Haas van Alphen quantum oscillations in single crystals of the topological semimetal {\ca} with torque magnetometry in high magnetic fields up to 35 T. In conjunction with density functional theory based calculations, the observed quantum oscillations frequencies indicate that the Fermi surfaces of {\ca} enclose an odd number of time-reversal-invariant momenta, satisfying one of the proposed criteria to realize topological superconductivity. Nonzero Berry phases extracted from the magnetic oscillations also support the nontrivial topological nature of {\ca}.}

\end{abstract}

\date{\today}
\maketitle

%% introduction
%%% TSC

Topological superconductors (TSCs) hosting Majorana fermions on the boundaries have recently attracted much attention because of potential application in quantum computing and other areas [1]. Although extensive studies to explore TSCs, such as metal-intercalated Bi$_{2}$Se$_{3}$ \cite{hor10a,krien11,liu15,shrut15} and half Heusler systems \cite{butch11,nakaj15a}, have been conducted, unambiguous experimental identification of topological superconductivity is still lacking.

%%% Fu theory
According to a theory proposed by Fu and Berg \cite{fu10}, a time-reversal-invariant centrosymmetric superconductor is a topological superconductor, if it possesses the following properties: (1) odd-parity pairing symmetry with a full superconducting gap and (2) an odd number of time-reversal-invariant momenta (TRIM) in the Brillouin zone, enclosed by its Fermi surfaces. Together with the study of the superconducting gap structure in the superconducting state, detailed investigation of Felmiology in the normal state is crucial for identifying topological superconductivity in topological materials.

%%% CaSn3 
Recently, the binary stannide semimetal {\ca} has been proposed to be a promising candidate for realizing topological superconductivity, as it is predicted to be a topologically nontrivial semimetal \cite{gupta17}. The nontrivial electronic band structure harbors topological nodal lines in the absence of spin orbit coupling (SOC). Upon turning on SOC, the nodal lines evolve into topological point nodes \cite{gupta17}. More notably, superconductivity has experimentally been confirmed  \cite{luo15a,zhu19}, and the possible nontrivial Berry phase associated with the topological nature of {\ca} has been obtained from a recent quantum oscillation study in the normal state. However, it is still unclear that this system satisfies the proposed criterion for topological superconductors, i.e. an odd number of Fermi surfaces enclosing TRIM in the Brillouin zone.

Here we present a detailed study of de Haas van Alphen (dHvA) oscillations in {\ca} with torque magnetometry in high magnetic fields up to 35 T. We observe four fundamental dHvA frequencies associated with the Fermi surfaces as well as strong Zeeman splitting. With the aid of our band structure calculations, we assign the measured frequencies to the Fermi surfaces of {\ca}, unveiling that an odd number of TRIM is enclosed by the Fermi surfaces. Combined with the observation of nontrivial Berry phases, this finding suggests that {\ca} is a promising material to realize topological superconductivity.

%% methods
Single crystals of {\ca} were grown by a Sn self flux method. The starting elements with a ratio of Ca:Sn = 1:9 were placed in an alumina crucible, which in turn was sealed in a quartz tube. The mixture was heated up to 800$^{\circ}$C, kept for 24 hours, and cooled slowly down to 300$^{\circ}$C at a rate of 2$^{\circ}$C/h. The excess of Sn flux was decanted by centrifugation.
%% crystal structure 
We confirmed the cubic AuCu$_{3}$-type structure with space group $Pm\bar{3}m$ with the lattice constant of $a$ = 4.7331(5) {\AA} via powder x-ray diffraction. No Sn impurities were observed unlike the previous reports \cite{luo15a,zhu19}.
%% torque method
Torque magnetometry was performed using a capacitive cantilever in a 35 T resistive magnet at the National High Magnetic Field Laboratory (NHMFL), Tallahassee, FL.

%% theory
The band structure and Fermi surfaces are calculated by means of density functional theory (DFT) employing the projector-augmented wave (PAW) pseudopotential method \cite{RN424, RN449} and a plane-wave basis set as implemented in the Vienna Ab initio Simulation Package \cite{RN429, RN425}. Exchange correlation effects are included via the generalized-gradient approximation (GGA) in the form of Perdew-Berke-Enzerhoff (PBE) exchange-correlation functionals \cite{RN453, RN435}. We set a cutoff energy of 500 eV for plane-wave expansion, and adopt the DFT+U approach \cite{RN17910} to mitigate the lack of electron correlation in the PBE functionals. The effective Hubbard $U$ parameter is chosen to be 2.55 eV to produce a lattice parameter 4.741 {\AA}, close to a value of 4.742 {\AA} that was determined experimentally \cite{luo15a} and not far from our measured value, $a$ = 4.7331(5) {\AA}. All electronic iterations converged with 0.01 meV threshold. We use a Gaussian smearing of 0.1 eV and sample the Brillouin Zone with a 15×15×15 $\Gamma$-centered grid. SOC is incorporated in the calculations of electronic band structure of the system. We calculate eigenvalues for all $k$-points of a 41$\times$41$\times$41 uniform grid.

%%% Results

%% torque

\begin{figure}[t]
  \includegraphics[width=8cm]{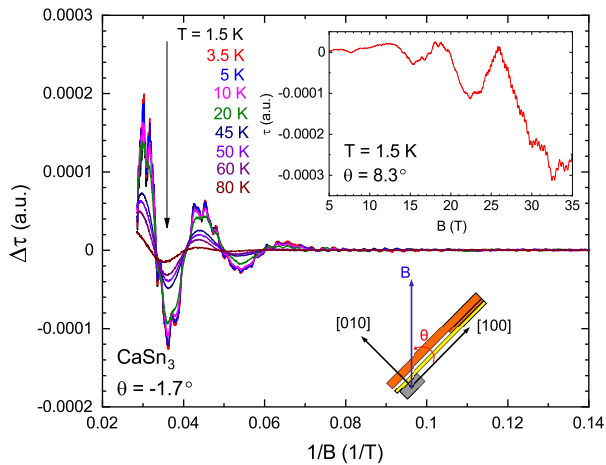}
\caption{{\bf Temperature dependent oscillatory magnetic torque in {\ca}}. Oscillatory part of magnetic torque as a function of $1/B$ at $\theta$ = -1.7$^{\circ}$, where $\theta$ is the angle between the magnetic field and the [100] axis while rotating the field from [100] to [010] in the (001) plane as illustrated in the lower inset. The oscillation amplitudes decrease with increasing temperatures. Upper inset: Magnetic torque as a function of the magnetic field $B$ for $\theta$ = 8.3$^{\circ}$ at $T$ = 1.5 K.}
\end{figure}

Our torque magnetometry data exhibit clear dHvA oscillations. Upon applying magnetic field at $\theta = 8.3^{\circ}$, where $\theta$ is the angle between the applied magnetic field and the [100] axis in the (001) plane, the oscillations in the magnetic torque $\tau$ are discernible above $\sim$ 5 T at 1.5 K (fig.1 inset). The magnitudes of the oscillatory part of the magnetic torque $\Delta \tau$ decrease rapidly with increasing temperature (fig.1). While the fast oscillations are completely suppressed above 20 K, the slow oscillations can be observed up to at least 80 K.

\begin{figure}[t]
\includegraphics[width=8cm]{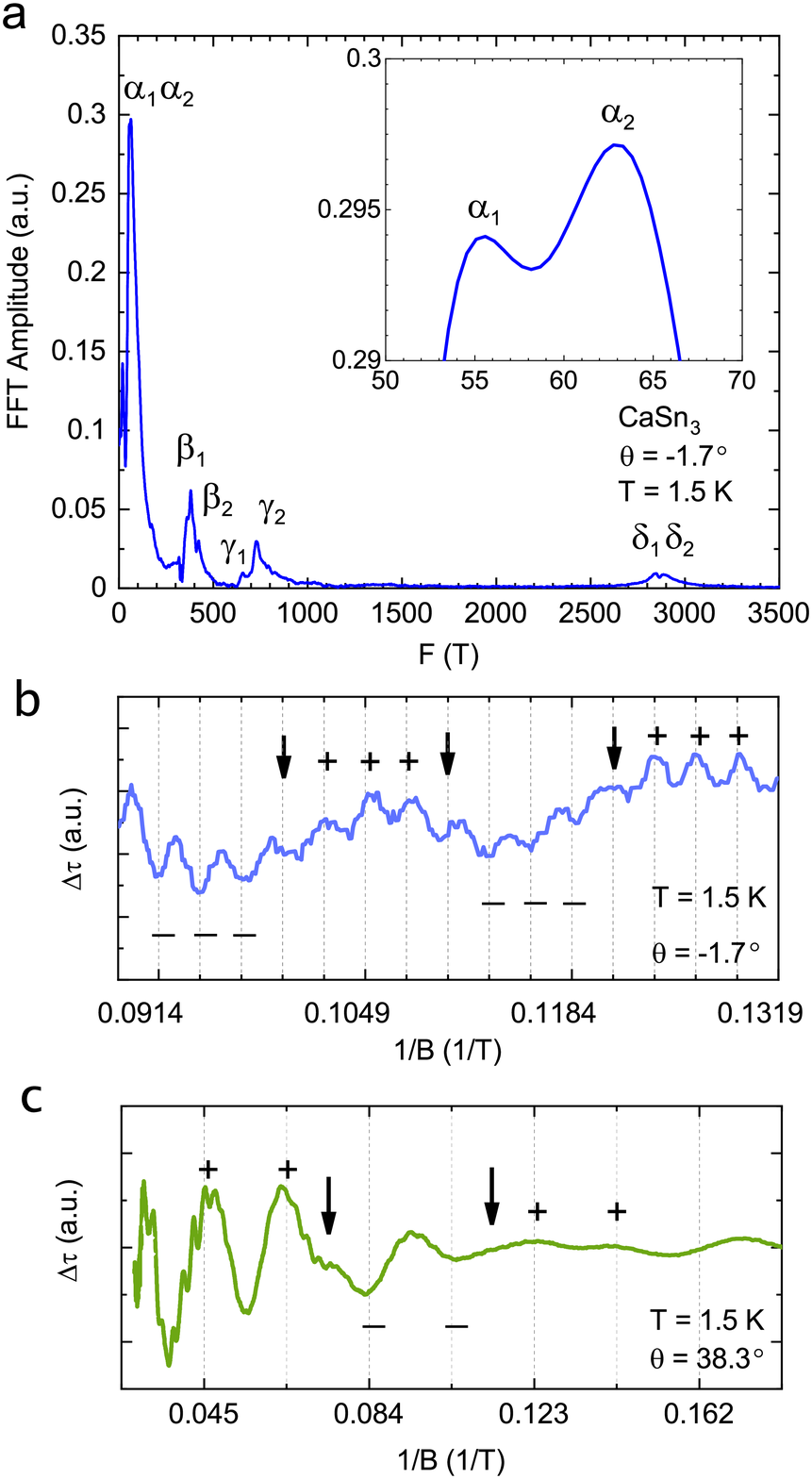}
\caption{{\bf Fast Fourier Transform (FFT) spectrum of oscillatory magnetic torque and peak splitting in {\ca}}. (a) FFT spectrum of oscillatory magnetic torque for $\theta$ = -1.7$^{\circ}$ at $T$ = 1.5 K (field range 5–35 T). Four fundamental frequencies, $F_{\alpha}$, $F_{\beta}$, $F_{\gamma}$, and $F_{\delta}$, with splittings are observed. Inset: Peak splitting in the $\alpha$ orbit at $T$ = 1.5 K. We observe beating due to two oscillatory components associated with (b) the $\beta$ orbit with the periodicity of 0.0027 T$^{-1}$ ($F\sim$ 370 T) for $\theta$ = -1.7$^{\circ}$ at $T$ = 1.5 K, and (c) the $\alpha$ orbit with the periodicity of 0.0195 T$^{-1}$ ($F\sim$ 50 T) for $\theta$ = 38.3$^{\circ}$ at $T$ = 1.5 K. Arrows indicate nodes in the beating patterns. Plus and minus signs indicate the positions of the maxima and minima of the oscillations, respectively.}
\end{figure}

Fast Fourier transform (FFT) allows us to determine the frequencies of oscillation components measured at $\theta$ = -1.7$^{\circ}$ (fig.~2a). We find four fundamental frequencies $F_{\alpha}$, $F_{\beta}$, $F_{\gamma}$, and $F_{\delta}$. Each of them in fact consists of two frequencies close to each other (fig.~2a and the inset), identified to be $F_{\alpha 1}$ = 56 T, $F_{\alpha 2}$ = 63 T, $F_{\beta 1}$ = 380 T, $F_{\beta 2}$ = 422 T, $F_{\gamma 1}$  = 655 T, $F_{\gamma 2}$ = 729 T, $F_{\delta 1}$ = 2845 T, and $F_{\delta 2}$ = 2890 T. This paring of close frequencies is corroborated by clear beating patterns in the magnetic torque as a function of $1/B$. As indicated by arrows in figs.~2b and c, nodes in the beating patterns produce $\pi$ phase shifts in the oscillations. While the frequencies of $F_{\alpha}$, $F_{\beta}$, and $F_{\gamma}$ are in good agreement with those previously reported\cite{zhu19}, $F_{\delta}$ that is prominent above 30 T (or $1/B <$ 0.033 T$^{-1}$) is newly identified by the present work.

The observed frequencies $F$ are associated with extremal cross sectional areas, $S$, of the Fermi surfaces in the momentum space by the Onsager relation $F=(\hbar/2\pi e)S$. Tracking the extremal cross sectional areas in rotating magnetic fields about the crystallographic axes, we can trace the Fermi surface topology of {\ca}. The angular dependence of dHvA oscillations in {\ca} about the [001] axis is plotted in fig.3a. As expected for the cubic structure of {\ca}, we observe symmetric behavior in the magnetic torque with respect to $\theta \sim 45^{\circ}$, similar to the previous report \cite{zhu19}. The dHvA frequencies extracted from the FFT analysis exhibit nearly isotropic angular dependence  for the $\alpha$, $\beta$, $\gamma$, and $\delta$ orbits, as shown in fig.3b.

%% band calculation

To compare with the experimental observations, we show the calculated band structure of {\ca} in fig.4a. The predicted bands consist of a hole band (band 25) and an electron band (band 26). Band 25 comprises a large hole pocket around the $\Gamma$ point in the Brillouin zone, surrounded by numerous small hole pockets (fig.4b), and band 26 has cross-shaped electron pockets around the X points and triangle-shaped electron pockets located between the $\Gamma$ and $R$ points, along with tiny electron pockets around the M points (fig.4c). Figure 3c shows theoretical quantum oscillation frequencies generated from the DFT calculations using the supercell K-space extremal area finder (SKEAF) code \cite{rourk12}.

\begin{figure}[t]
  \includegraphics[width=8cm]{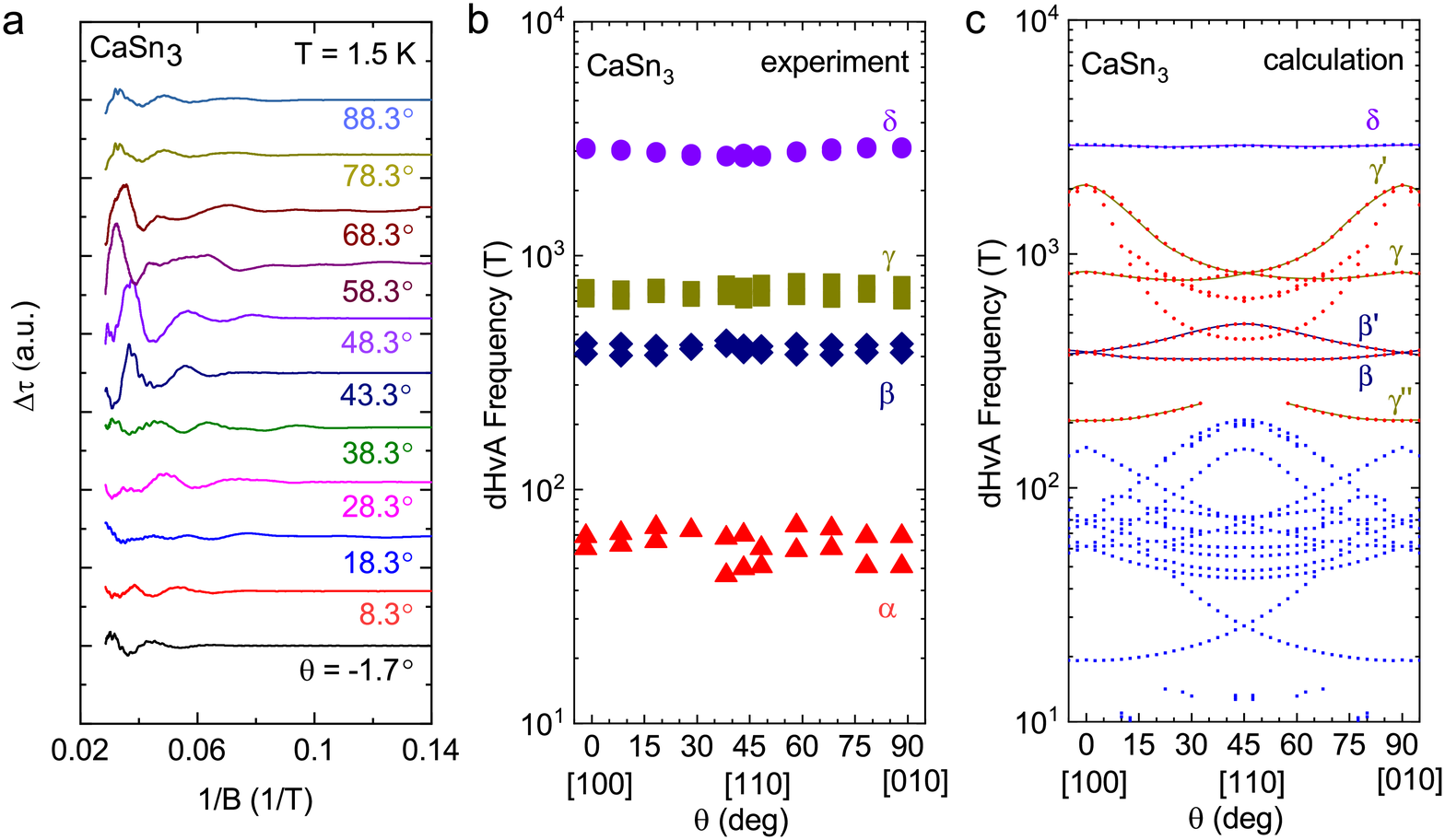}
\caption{{\bf Angular dependence of dHvA oscillation frequencies.} (a) Angular dependence of the oscillatory part of magnetic torque as a function of $1/B$. Reflecting the cubic crystal structure, the oscillations are symmetric with respect to $\theta\sim$ 45$^{\circ}$. (b) Angular dependence of dHvA oscillation frequencies. The observed frequencies $F_{\alpha}$, $F_{\beta}$, $F_{\gamma}$, and $F_{\delta}$ are nearly independent of angle $\theta$. (c) Theoretical results obtained from DFT calculations using the SKEAF code for band 25 (blue) and band 26 (red). }
\end{figure}

\begin{figure}[t]
\includegraphics[width=8cm]{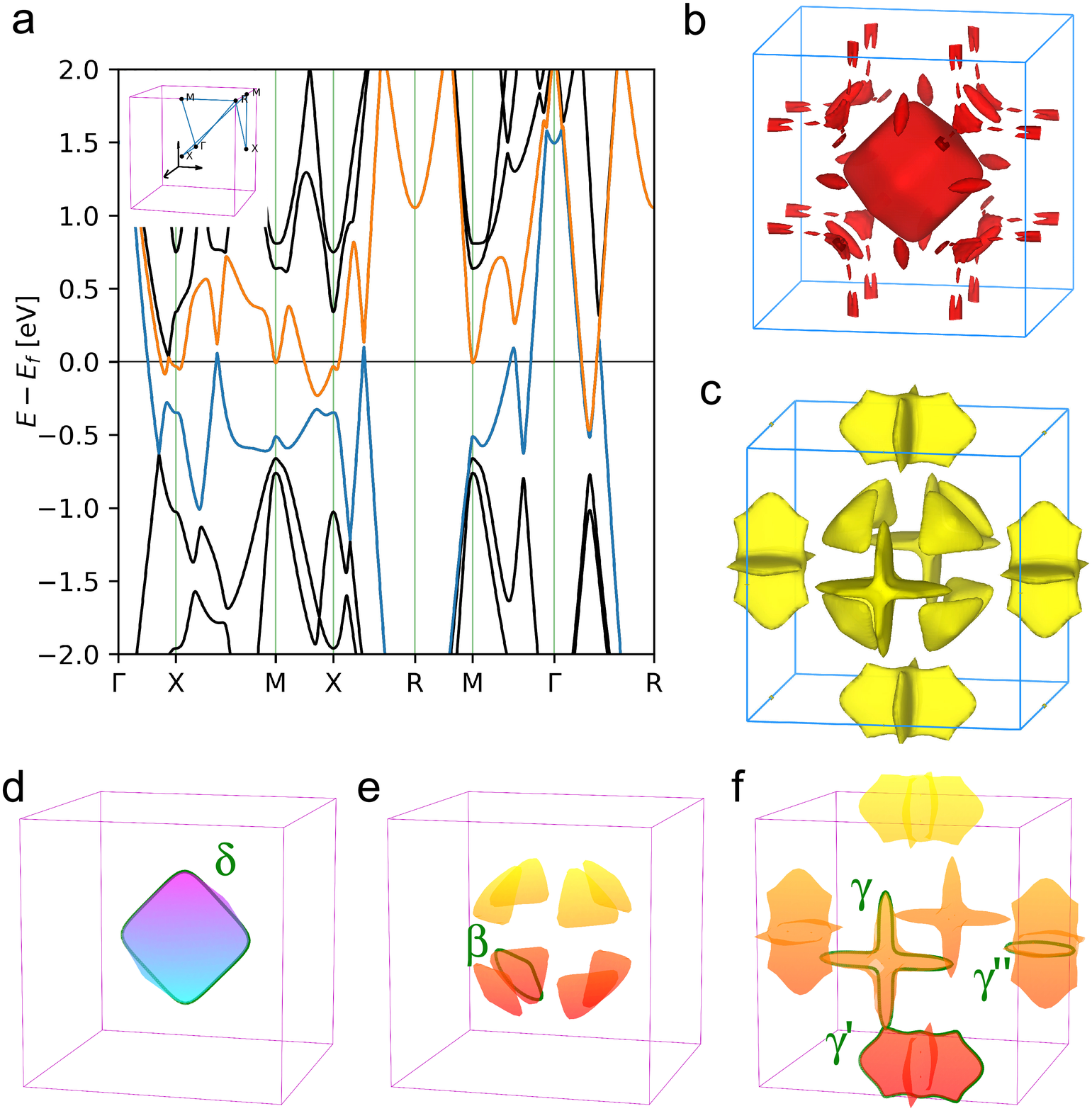}
\caption{{\bf Electronic band structure and Fermi surfaces of {\ca}}. (a) Band structure of {\ca} calculated by DFT. The local minimum of the electron band (shown in orange) at the M point is very close to the Fermi energy $E_{F}$, suggesting that the sizes of Fermi pockets at the M points are extremely sensitive to $E_{F}$. Inset: high symmetry points in the Brillouin zone. (b) Fermi surfaces of band 25 (hole). A large hole pocket is located around the $\Gamma$ point, surrounded by numerous tiny hole pockets. (c) Fermi surfaces of band 26 (electron), consisting of cross-shaped electron pockets around the X points, triangle-shaped electron pockets between the $\Gamma$ and R points, and small electron pockets around the M points. Calculated dHvA orbits for (d) $\delta$, (e) $\beta$, (f) $\gamma, \gamma^{\prime}$, and $ \gamma^{\prime\prime}$. Corresponding frequencies are plotted in fig.3c.}
\end{figure}

On comparing the theoretical oscillation frequencies with the observed ones, we can allocate the frequencies to corresponding Fermi surfaces of {\ca}. The $\delta$ orbit with $F_{\delta}\sim$ 2900 T corresponds well to the large hole pocket around the $\Gamma$ point (fig.4d). The $\beta$ orbit with $F_{\beta}\sim$ 400 T is close to the frequency branch of triangle-shaped electron pockets as shown in fig.4e, and the $\gamma$ orbit with $F_{\gamma}\sim$ 690 T is comparable to the frequency associated with the cross-shaped electron pockets around the X points (fig.4f).

However, there are important discrepancies between the theoretical results and the measured dHvA frequencies. We observe nearly isotropic angular dependence of dHvA frequencies, and the splittings of frequencies are also isotropic. In the $\delta$ orbit, the observed splitting is $\Delta F_{\delta}^{obs}\sim 45$ T, but the theoretical calculations predict no splitting in the branch near $\sim$2900 T. In the $\beta$ and $\gamma$ orbits, the observed splittings are $\Delta F_{\beta}^{obs} \sim$ 40 T and $\Delta F_{\delta}^{obs} \sim$ 70 T. However, the theoretical calculations indicate that the frequencies for the $\beta^{\prime}$ and $\gamma^{\prime}$ orbits, stemming from the triangle-shaped and cross-shaped electron pockets (figs.4e and f), respectively, depend strongly on the angle $\theta$ as shown in fig.3c, yielding angular dependent splitting in the oscillation frequencies. Moreover, the splittings of calculated frequencies $\Delta F_{\beta}^{calc} = F_{\beta^{\prime}}^{calc}-F_{\beta}^{calc}$ = 150 T at $\theta = 45^{\circ}$ and $\Delta F_{\gamma}^{calc}= F_{\gamma^{\prime}}^{calc}-F_{\gamma}^{calc}$ = 1200 T at $\theta = 0^{\circ}$ and 90$^{\circ}$ are much larger than the observed $\Delta F_{\beta}^{obs}$ and $\Delta F_{\gamma}^{obs}$. We therefore conclude that the experimentally detected $\beta$ and $\gamma$ orbits represent only parts of the corresponding theoretical orbits. The absence of these angle-dependent branches in our measurements can be attributed to the suppression of oscillation amplitudes by the large curvature factor $\Delta \tau \propto |\partial^{2} S(k)/\partial k_{\parallel}^{2}|^{-1/2}$, suggested by the shape of calculated Fermi surfaces.

\begin{figure}[t]
\includegraphics[width=8cm]{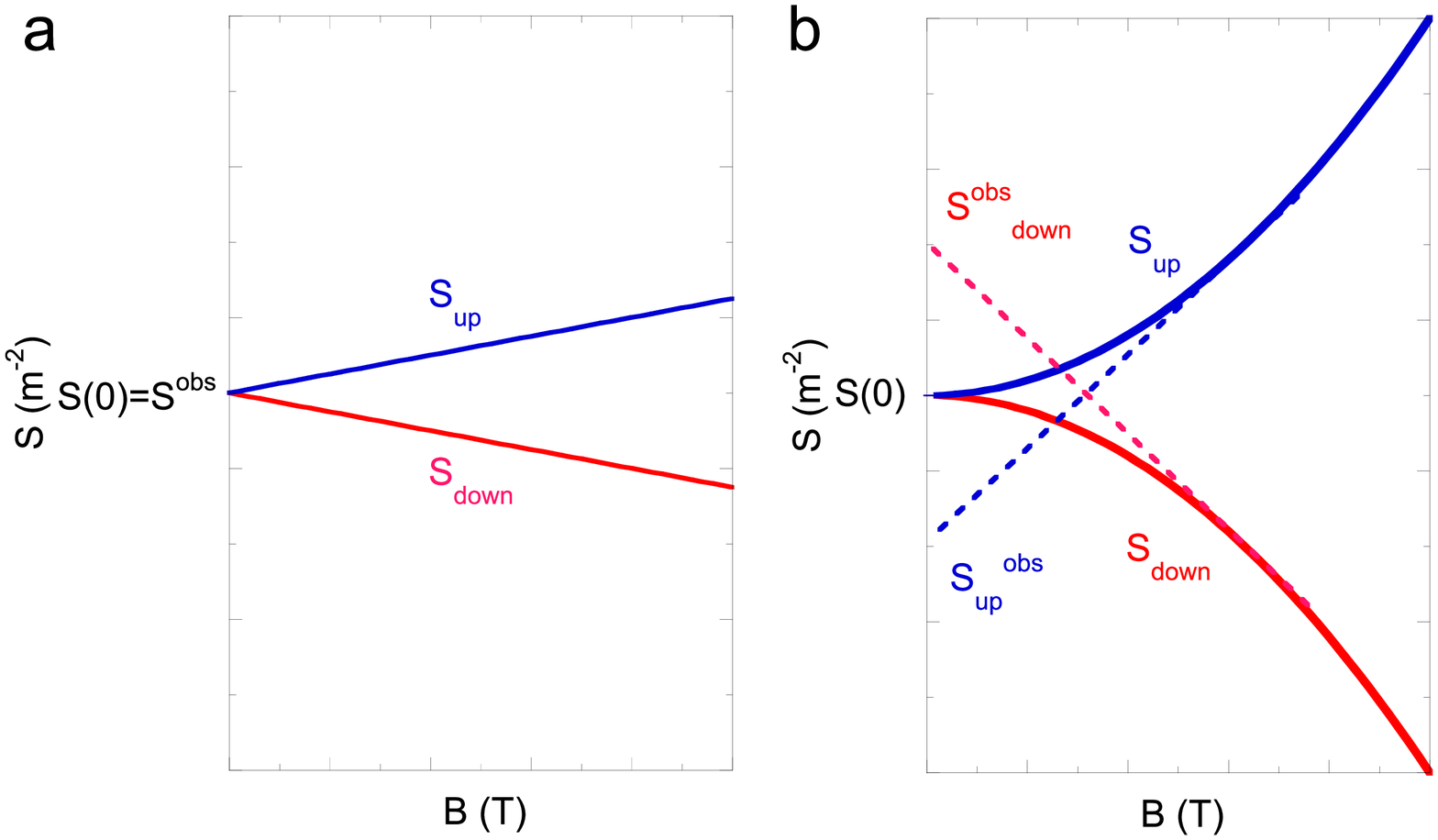}
\caption{{\bf Spin-splitting of Fermi surface due to nonlinear Zeeman effect.} (a) Magnetic field dependence of the extremal cross-sectional areas for spin-up and down Fermi surfaces $S_{up}$ and $S_{down}$ due to the linear Zeeman effect. The observed extremal cross-sectional area $S^{obs}=S(B)-BdS(B)/dB$ is identical to the zero-field limit of cross-sectional area $S(0)$, leading to $S_{up}^{obs}=S_{down}^{obs}$. (b) Magnetic field dependence of the extremal cross-sectional areas for spin-up and spin-down Fermi surfaces due to the nonlinear Zeeman effect. The observed extremal cross-sectional area is given by the extrapolation of $dS(B)/dB$ to $B = 0$, yielding $S_{up}^{obs}\neq S_{down}^{obs}$.}
\end{figure}

\begin{figure*}[tb]
\includegraphics[width=15cm]{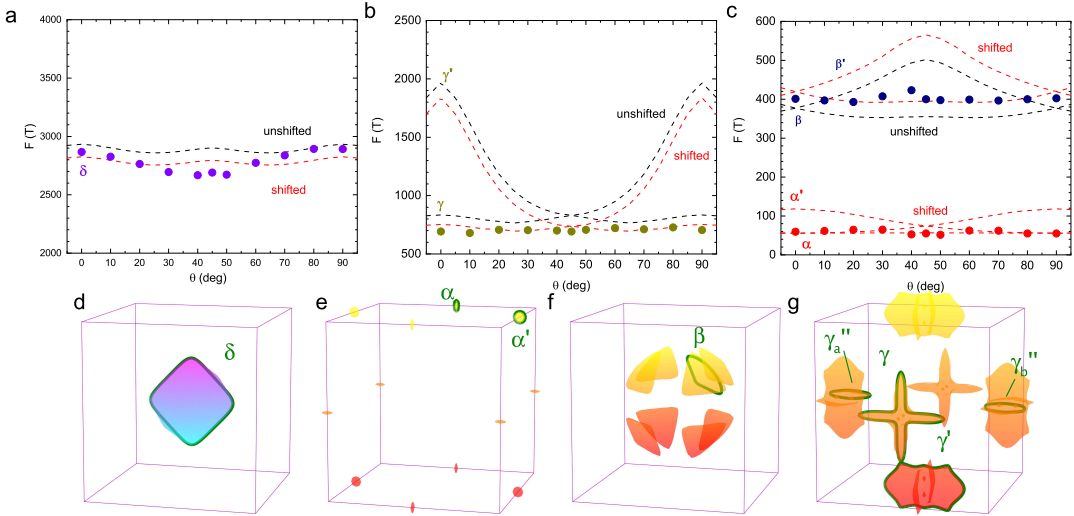}
\caption{{\bf Comparison between experimental and calculated dHvA frequencies.} Angular dependence of average dHvA frequencies for (a) the $\delta$ orbit, (b) the $\gamma$ orbit, and (c) the $\alpha$ and $\beta$ orbits. Black and red dashed lines represent theoretical dHvA frequencies without and with Fermi-energy shifts, respectively. Fermi surfaces with  shifting the Fermi energy by (d) +40 meV for the hole band aound the $\Gamma$ point, (e) +110 meV for electron pockets around the M points, (f) +15 meV for electron pockets located between the R and $\Gamma$ points, and -20 meV for electron pockets around the X points. Green lines represent extremal orbits when the field is applied parallel to the [100] axis.}
\end{figure*}

The slight splitting of oscillation frequencies observed in {\ca} is reminiscent of the spin splitting of energy bands. For instance, in noncentrosymmetric systems with strong SOC, asymmetric SOC can induce the splitting of Fermi surfaces even in the absence of a magnetic field \cite{minee05,butch11}. However, since {\ca} is a centrosymmetric system, asymmetric SOC is absent. Instead, the observed splitting in {\ca} can be attributed to spin-up and down Fermi surfaces with field-dependent extremal cross sectional areas $S(B)$ due to the Zeeman effect. Taking the Zeeman effect into account, the Onsager relation can be rewritten as $F= \hbar(S(B)-BdS(B)/dB)/(2\pi e)$. Hence, the measured $F$ represents a back projection of field-dependent cross-sectional area $S(B)$, i.e. an extrapolation of the tangent at $B$ to $B=0$ (fig.5) \cite{shoen11,mccol05,rourk08,mercu09}. In the conventional linear Zeeman effect, the back projection provides field-independent $S^{obs}=S(0)$, or field-independent dHvA frequencies $F^{obs}$ (fig.5a). In this case, the observed frequencies for spin-up and down Fermi surfaces are identical $(F^{obs}_{up}=F^{obs}_{down})$. On the other hand, the nonlinear Zeeman effect can give rise to nonlinear magnetic field dependence of extremal areas $S_{up}(B)$ and $S_{down}(B)$, yielding the disparity between the observed extremal areas $S_{up}^{obs}$ and $S_{down}^{obs}$, i.e. $F^{obs}_{up} \neq F^{obs}_{down}$ (fig.5b). In this situation, averaging two splitting frequencies provides the zero-field limit for each orbit: $F_{\alpha}^{ave}$ = 60 T, $F_{\beta}^{ave}$ = 401 T, $F_{\gamma}^{ave}$ = 692 T, and $F_{\delta}^{ave}$ = 2868 T. Shifting the Fermi energy upward by 15 meV for the $\beta$ orbit, downward by 20 meV for the $\gamma$ and $\gamma^{\prime}$ orbits, and upward by 40 meV for the $\delta$ orbit fully reproduce their absolute frequencies (figs.6a-c). Field-dependent dHvA frequencies for spin-up and down Fermi surfaces have also been reported for PrPb$_{3}$ with the same AuCu$_{3}$-type cubic structure \cite{endo02}.

%% M pocket

Whereas the Fermi pockets corresponding to the measured orbits $\beta$, $\gamma$, and $\delta$ are uniquely identified, the $\alpha$ orbit with $F_{\alpha}^{ave}= 60$ T remains unassigned. We note that the hole pockets around the $\Gamma$ points and triangle-shaped electron pockets between $\Gamma$ and R points provide no extremal orbits other than $F\sim 3000$ T and $\sim 400$ T at $\theta$ = 0, respectively (figs.4d, 4e, 6d and 6f). On the other hand, the cross-shaped electron pockets around the X points has extra extremal orbits $\gamma^{\prime\prime}$ with the frequency of $\sim$ 190 T (150 T) for the unshifted (shifted) Fermi pockets (figs.4f and 6g) at $\theta$ = 0. Since the calculated frequencies for the $\gamma^{\prime\prime}$ orbit, with or without the Fermi-energy shift, are larger by a factor of $\sim$3 than the measured frequency of 60 T, the $\alpha$ orbit cannot be ascribed to the cross-shaped electron pockets. Moreover, the isotropy of the dHvA frequencies for the $\alpha$ orbit suggests that it is quite unlikely that the $\alpha$ orbit originates from the small nonspherical hole pockets located off the high symmetry points in band 25.

We therefore attribute the $\alpha$ orbit to the electron pockets around the M points. In the band structure with the unshifted Fermi energy, however, the cross sectional areas of the electron pockets around the M points are very small (fig.4c). Nevertheless, as shown in the calculated band structure (fig.4a), the sizes of the electron pockets are extremely sensitive to the Fermi energy $E_{F}$. Indeed, an upward shift of $E_{F}$ by 110 meV yields perfect agreement in the absolute frequency and its angular dependence (fig.6c) just like the other Fermi pockets in {\ca}. This assignment to the $\alpha$ orbit, with a shifted Fermi energy, leads to the intriguing conclusion that the three Fermi pockets in {\ca} surround the TRIM---namely, the $\Gamma$, X, and M points---satisfying one of the criteria theoretically proposed for the realization of topological superconductivity.

%% effective mass

\begin{figure}[thb]
\includegraphics[width=9cm]{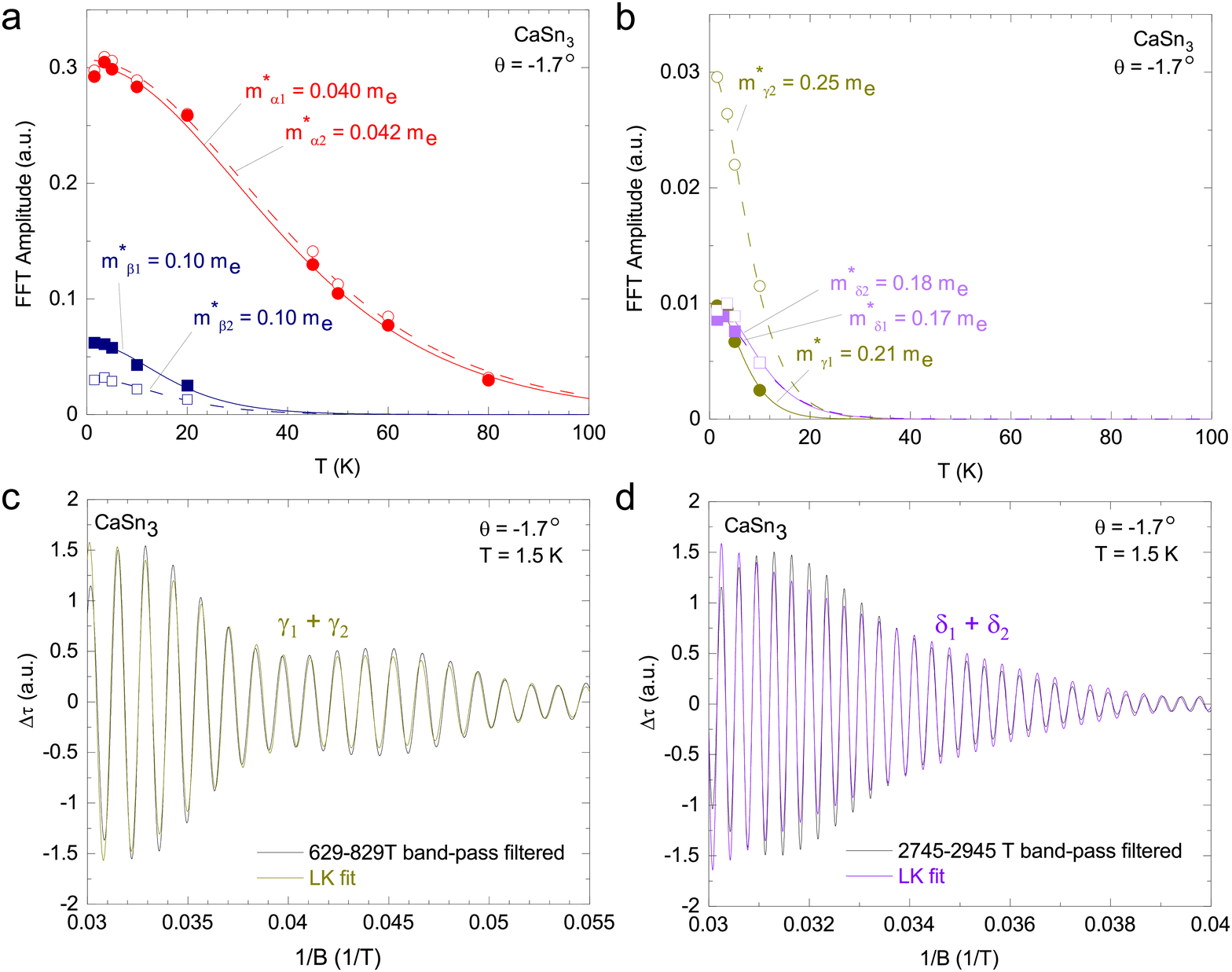}
\caption{{\bf Effective masses and Berry phases of {\ca}}. Effective masses of (a) the $\alpha$ and $\beta$ orbits and (b) the $\gamma$ and $\delta$ orbits in {\ca}, obtained from temperature dependence of FFT amplitudes at $\theta$ = -1.7$^{\circ}$ (fig.2a) using the Lifshitz-Kosevich model with the average inverse field $1/{\bar B}=(1/B_{min}+1/B_{max})/2$, where $B_{min}$ = 5 T and $B_{max}$ = 35 T. A two component LK fit to the band-pass filtered data for (c) the $\gamma$ and (d) the $\delta$ orbits. The extracted Berry phase $\phi_{B}$---0.7$\pi$ for the $\gamma$ orbit and 1.0$\pi$ for the$\delta$ orbit---indicate nontrivial topological nature of {\ca}.}
\end{figure}

To take a closer look at the effective mass and topological nature of each band, we utilize the Lifshitz-Kosevich (LK) model that explains the oscillatory part of magnetic torque in metals \cite{shoen11},
\begin{align}
\Delta \tau^{osc} = M_{\perp}^{osc}\times B,
\end{align}
\begin{align}
M_{\perp}^{osc} = -\frac{1}{F}\frac{dF}{d\theta}M_{\parallel}^{osc},
\end{align}
\begin{align}
M_{\parallel}^{osc}\propto -{B}^ {\frac{1}{2}}\left |\frac{\partial^{2} S(k)}{\partial k_{\parallel}^{2}}\right |^{-\frac{1}{2}} R_{T}R_{D}\sin\left[2\pi \left (\frac{F}{B}+\phi^{\sigma}\right )\right].
\end{align}
Here, $R_{T}= x/\sinh x$---with $x = a m^{\ast}T/m_{e}B$ and $a = 2\pi^{2}k_{B}m_{e}/e\hbar = 14.69$ T/K---is the thermal damping factor, $R_{D} = \exp(-a m^{\ast}T_{D}/B)$ is the Dingle damping factor, $m^{\ast}$ is the carrier effective mass, $m_{e}$ is the free electron mass, and $T_{D}$ is the Dingle temperature. The phase shift $\phi^{\sigma}$ ($\sigma$ = up, down) for a spin-up/down Fermi surface is given by $\phi^{\sigma} = -1/2 + \phi_{B}^{\sigma}/2\pi \pm \phi_{Z}^{\sigma}/2 + \phi_{3D}$, where $\phi_{B}^{\sigma}$ is the Berry phase and $\phi_{Z}^{\sigma} = gm^{\ast}/2m_{e} $ ($g$: g-factor) is a phase shift due to the linear Zeeman effect. The phase shift determined by dimensionality of the Fermi surface, $\phi_{3D}$, is 1/8 for extreme minima (maxima) for electron (hole) pockets and -1/8 for extreme maxima (minima) for electron (hole) pockets. Note here that instead of using the spin damping factor $R_{S}=\cos(\pi gm^{\ast}/2m_{e})$ in the conventional LK formula, we adopt the Zeeman phase shift $\phi_{Z}^{\sigma}$ to describe the effect of spins due to the spin-dependent dHvA frequencies $(F_{up}\neq F_{down})$. To extract a reliable Berry phase from dHvA oscillations, $\phi_{Z}^{\sigma}$, together with the sign of $dF/d\theta$, should be considered in the LK analysis.

To this goal, we first determine the effective masses $m^{\ast}$ through the thermal damping factor $R_{T}$ by fitting the observed FFT amplitudes to the LK formula \cite{shoen11}. We find the effective masses $m^{\ast}$ are $0.040m_{e}$, $0.042m_{e}$, $0.10m_{e}$, $0.10m_{e}$, $0.21m_{e}$, $0.25m_{e}$, $0.17m_{e}$, and $0.18m_{e}$ for the $\alpha_{1}$, $\alpha_{2}$, $\beta_{1}$, $\beta_{2}$, $\gamma_{1}$, $\gamma_{2}$, $\delta_{1}$, and $\delta_{2}$ orbits, respectively (figs.7a and b). These light masses are consistent with the previously reported values \cite{zhu19}.

%% phase shift

Nontrivial topological nature of {\ca} is confirmed by the nonzero Berry phase. To determine the Berry phase, we fit a two-component LK formula to band-pass filtered data for the $\gamma$ and $\delta$ orbits for $\theta$ = -1.7$^{\circ}$ at $T$ = 1.5 K. We use the oscillation frequencies obtained from the FFT analysis and the effective masses extracted from the temperature dependence of FFT amplitudes as fixed parameters. We assume that the spin-up and down Fermi surfaces have the same Berry phase, i.e. $\phi_{B}=\phi_{B}^{up}=\phi_{B}^{down}$. Taking into account the signs of $dF/d\theta$ ($<0$ around $\theta = 0$) for the $\gamma$ and $\delta$ orbits (figs.6a and b), we find the extracted Berry phases $\phi_{B} = 0.7\pi$ for the $\gamma$ orbit and $\phi_{B} = 1.0\pi$ for the $\delta$ orbit, which indicate the nontrivial topological nature of these bands. The lower limits of effective $g$-factors, obtained from the Zeeman phase shift, are 8.61 for the $\gamma$ orbit and 11.6 for the $\delta$ orbit. Similar, sizable enhancement of $g$ has been observed in various topological semimetals, including ZrSiS \cite{hu17}, ZrTe$_{5}$ \cite{liu16}, Cd$_{3}$As$_{2}$ \cite{cao15}, and PtBi$_{2-x}$ \cite{xing20}, all consistent with the large Zeeman splitting.

In summary, we have observed four fundamental dHvA oscillation frequencies in {\ca} via torque magnetometry in magnetic fields up to 35 T, determined to be $F_{\alpha}^{ave}$ = 60 T, $F_{\beta}^{ave}$ = 401 T, $F_{\gamma}^{ave}$ = 692 T, and $F_{\delta}^{ave}$ = 2868 T. We have identified the correspondence between the experimental quantum oscillation frequencies and theoretically calculated orbits in this topological superconductor candidate, revealing that an odd number of Fermi pockets enclose the TRIM, prerequisite to one of the theoretical criteria for topological superconductivity. The nonzero Berry phases are also confirmed by using the LK model, supporting the nontrivial topological nature of this system. These findings provide a new avenue to investigate topological superconducting states stabilized in topological semimetals.

\begin{acknowledgments}
  The authors thank H. Shishido for helpful discussions. The experimental work was supported in part by the start-up fund from the University of Central Florida. H.S and Y.N. were supported by NSF CAREER DMR-1944975, and X.H. and Y.T. by the NHMFL UCGP program. Theoretical work (D.L. and T.S.R,) was supported by DOE Grant DE‐FG02‐07ER46354. The NHMFL is supported by the National Science Foundation through NSF/DMR-1644779 and the State of Florida.
\end{acknowledgments}

$*$ Correspondence and requests for materials should be addressed to Y.N. (Email: Yasuyuki.Nakajima@ucf.edu).

\end{document}